\documentclass[aps,pre,reprint,showpacs,notitlepage,onecolumn,superscriptaddress,numbers]{revtex4-1}
\usepackage{epsfig, graphics,amssymb,amsmath,subeqnarray, color, bm,mathtools,xcolor, float,graphicx}
\usepackage[toc,page]{appendix}
\usepackage{commath}
\usepackage{caption}
\usepackage{subcaption}
\numberwithin{equation}{section}

\providecommand{\bnabla}{\boldsymbol{\nabla}}
\providecommand{\bcdot}{\boldsymbol{\cdot}}
\providecommand{\mathsfbi}[1]{\boldsymbol{\mathsf{#1}}}

\newcommand{\dGamma}{\dif\Gamma(\xibold)}
\newcommand{\GammaD}{\Gamma_{D}}
\newcommand{\GammaP}{\Gamma_{P}}

\newcommand{\chibold}{\boldsymbol{\chi}}
\newcommand{\xibold}{\boldsymbol{\xi}}
\newcommand{\rbold}{\boldsymbol{r}}

\newcommand{\vel}{\boldsymbol{u}}
\newcommand{\f}{\boldsymbol{f}}
\newcommand{\unitNormal}{\boldsymbol{n}}

\newcommand{\uD}{\vel^{D}}
\newcommand{\uP}{\vel^{P}}

\newcommand{\fD}{\f^{D}}
\newcommand{\fP}{\f^{P}}

\newcommand{\Gbold}{\mathsfbi{G}}
\newcommand{\Kbold}{\mathsfbi{K}}

\newcommand{\btau}{\mathsfbi{\boldsymbol{\tau}}}
\newcommand{\btaus}{\mathsfbi{\boldsymbol{\tau}}_s}

\begin{document}

\title{Effects of surface viscosities on the motion of a droplet enclosing a translating particle}

\author{Ali G\"urb\"uz}
\thanks{These authors contributed equally to this work.}
\affiliation{Department of Mechanical Engineering, Santa Clara University, Santa Clara, CA, USA}
\affiliation{Department of Mathematics, Towson University, Towson, MD 21252, USA}

\author{Herv\'e Nganguia}
\thanks{These authors contributed equally to this work.}
\affiliation{Department of Mathematics, Towson University, Towson, MD 21252, USA}

\author{Guangpu Zhu}
\affiliation{Department of Mechanical Engineering, National University of Singapore, Singapore}
\affiliation{College of Aerospace Engineering, Nanjing University of Aeronautics and Astronautics, Nanjing 210016, China}

\author{Lailai Zhu}
\affiliation{Department of Mechanical Engineering, National University of Singapore, Singapore}

\author{Y. N. Young}
\affiliation{Department of Mathematical Sciences, New Jersey Institute of Technology, Newark, NJ 07102, USA}

\author{On Shun Pak}
\email{Electronic mail: opak@scu.edu}
\affiliation{Department of Mechanical Engineering, Santa Clara University, Santa Clara, CA, USA}

\date{\today}

\begin{abstract}
We investigate the influence of interfacial rheology on the motion of a compound particle consisting of a viscous droplet enclosing a translating rigid particle in the Stokes flow regime. The droplet interface is modeled using the Boussinesq–Scriven constitutive law, incorporating both surface shear and dilatational viscosities. An exact analytical solution is derived for the concentric configuration, and the analysis is extended to eccentric geometries using a spectral boundary integral method, enabling a systematic examination of confinement, viscosity contrast, and interfacial properties. For concentric configurations, we show that the induced droplet velocity is independent of surface shear viscosity, while surface dilatational viscosity can either enhance or suppress the droplet motion depending on the interplay between confinement and viscosity ratio. This behavior is rationalized in terms of competing effects between reduced interfacial mobility and increased driving force required to maintain the prescribed particle speed. In contrast, when the particle is eccentrically positioned within the droplet, a dependence on surface shear viscosity emerges, leading to a consistent enhancement of droplet motion that becomes more pronounced with increasing eccentricity. The analytical and numerical results are in excellent agreement and reveal how interfacial rheology, confinement, and symmetry breaking jointly govern the dynamics of compound particle systems. These findings provide mechanistic insight and establish a quantitative benchmark for future studies of active compound particles with complex interfaces.
\end{abstract}

\maketitle

\section{Introduction}
\label{Sec:Introduction}
Liquid drops enclosing various inclusions, sometimes referred to as compound particles, are ubiquitous in natural systems, for example, in dew or raindrops that encapsulate minerals, dust, and pollutants, or in respiratory droplets that carry pathogens \citep{Wisdom2013, Bozic2021, Pohlker2023}. In engineered systems, the encapsulation of diverse types of particles, such as biological cells, magnetic particles, and microswimmers \citep{Abate2009, Kemna2012, Brouzes2015, Ding2016, Ramos2020, Rajabi2021, Kokot2022}, has opened pathways toward a variety of microfluidic and biomedical applications \citep{He2005, Griffiths2006, Kelly2007, Chabert2008, Tormos2008, Thiele2011, Wen2015}. In particular, the ability to propel and manipulate individual droplets in a controlled manner represents a key operation in microfluidic applications. For example, \citet{Brouzes2015} demonstrated the encapsulation of superparamagnetic microbeads to extract target molecules from droplets in segmented flows, while \citet{Ding2016} explored the use of externally actuated magnetic helical micropellers to manipulate microdroplets.

The possibility of driving droplet motion through enclosed active particles or externally actuated inclusions has motivated fundamental investigations into the hydrodynamics of these complex systems, where the hydrodynamic interaction between the inclusion and the confining fluid interface gives rise to their rich dynamics \citep{Johnson1985, Sadhal1985, DaddiMoussaIder2018elastic}. Previous studies have examined the dynamics of compound particles driven internally by squirmers \citep{Reigh2017a,shaik_vasani_ardekani_2018, Kree2021_EPJE, Kree2022, Kree2023, Nganguia2025}, flow singularities \citep{Sprenger2020, Kree2021, Kawakami2025B, Kawakami_Vlahovska_2025}, as well as externally forced spherical particles \citep{KVS2019, Singeetham2021, Kree2021_EPJE}. While earlier studies primarily examined the dynamics of non-deforming spherical droplets, recent work \citep{Kawakami2025B, Kawakami_Vlahovska_2025} has revealed intriguing behaviours in deformable droplets containing active particles. 

Another line of research focuses on how complex interfacial behaviours influence the motion of compound particles. These investigations are motivated by both natural and engineered systems in which fluid interfaces are often laden with molecules or particles, giving rise to interfacial phenomena that cannot be described by a single scalar value of surface tension \citep{Fuller2012, Samaniuk2014, Manikantan2020, Jaensson2021}. For instance, prior studies have elucidated how Marangoni stresses arising from inhomogeneous surfactant distributions along the interface affect the dynamics of drops \citep{Stone1990, Vlahovska2009} and compound particle systems \citep{Mandal2016, shaik_vasani_ardekani_2018, ChembaiGanesh2023, Kawakami_Vlahovska_2025}. Beyond Marangoni effects, the transport of contaminants along the interface can generate shear and dilatational friction, introducing resistance to interfacial deformation characterized by surface viscosities \citep{Oldroyd1955, Fuller2012, Samaniuk2014, Langevin2014, Manikantan2016, Manikantan2020, Jaensson2021}. Earlier studies quantified the individual influences of surface shear and dilatational viscosities on the translational motion of non-deformable spherical drops \citep{Flumerfelt1980, LEVAN198111, Narsimhan2018}. More recent studies have extended these investigations to examine the motion of drops with surface viscosities under various background flows, as well as their deformation and breakup behaviours \citep{Rallison1984, Pozrikidis1994, Gounley_Boedec_Jaeger_Leonetti_2016, PonceTorres2017, Narsimhan2019, Dandekar_Ardekani_2020, Singh2020, Panigrahi2021, Herrada2022, Singh_Narsimhan_2022}. However, the influence of interfacial rheology on the dynamics of compound particles has only recently begun to be examined. \citet{Sprenger2020} presented an initial analysis of how surface shear viscosity affects the motion of droplets containing internal flow singularities, while \citet{Nganguia2025} later examined the motion of a squirmer concentrically enclosed within a spherical drop endowed with surface viscosities.

In this work, we investigate how interfacial viscous stresses influence the motion of a compound particle consisting of a non-deformable spherical drop that encloses an externally forced spherical inclusion. The same geometric configuration was previously examined by \citet{KVS2019} for a clean interface in the concentric case, and the present study extends their analysis by elucidating how the individual and combined effects of surface shear and dilatational viscosities alter the dynamics of such compound particle systems. We first present a theoretical analysis for the concentric configuration and subsequently employ simulations based on the boundary integral method to examine eccentric cases. Our findings address fundamental questions of how interfacial rheological effects enhance or hinder the motion of a drop internally driven by its inclusion, thereby providing mechanistic insights into the dynamics of such compound particle systems. The exact analytical solution and the numerical tools developed herein provide validation and benchmarks for future studies involving more complex configurations and interfacial behaviours.

\section{Problem formulation}
\label{Sec:ProblemFormulation}

\begin{figure}
    \centering
    \includegraphics[width=0.53\linewidth]{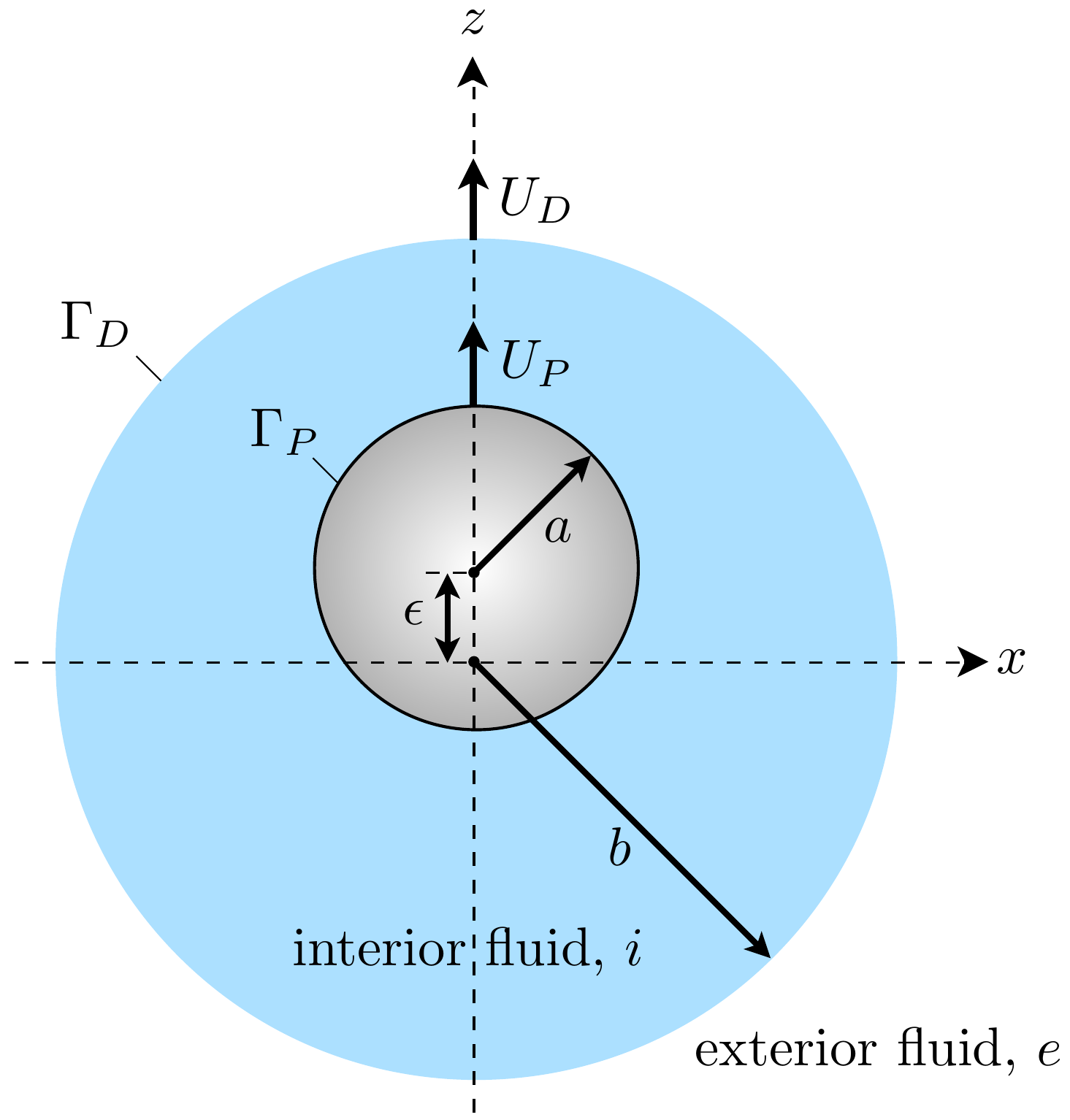}
\caption{Geometric setup and notation for the compound particle system. A spherical particle of radius $a$, with surface $\Gamma_P$, is enclosed within a spherical droplet of radius $b$, with interface $\Gamma_D$. The particle is driven at a prescribed speed $U_P$ along the $z$--axis, inducing a translational motion of the droplet at an unknown speed $U_D$ in the same direction by symmetry. The center of the particle lies at a distance from the droplet center defined as the eccentricity, $\epsilon$. The fluids inside and outside the droplet are denoted by $i$ and $e$, respectively.}
    \label{fig:SchematicForCompoundDropletProblem}
\end{figure}

We consider a compound particle system consisting of a rigid spherical particle of radius \(a\) enclosed within a spherical viscous droplet of radius \(b\), as illustrated in figure \ref{fig:SchematicForCompoundDropletProblem}. The distance between the centers of the particle and the droplet is defined as the eccentricity, $\epsilon$. The droplet is immersed in an unbounded Newtonian fluid. The interior region between the particle and the droplet interface is filled with a Newtonian fluid of viscosity $\mu_i$, while the exterior fluid has viscosity $\mu_e$. The droplet interface is denoted by $\Gamma_D$ and the particle surface by $\Gamma_P$. The rigid particle translates with a prescribed velocity $\boldsymbol{U}_P = U_P \boldsymbol{e}_z$, where $\boldsymbol{e}_z$ is the unit vector in the $z$--direction. This motion drives fluid flow both inside and outside the droplet and induces translation of the droplet itself with velocity $\boldsymbol{U}_D = U_D \boldsymbol{e}_z$, which is an unknown to be determined as part of the solution. We assume a vanishingly small Reynolds number such that the flow is governed by the Stokes equations in both fluid domains, namely,
\begin{equation}
	-\nabla p_j + \mu_j \nabla^2 \vel_j = \boldsymbol{0},
\end{equation}
\begin{equation}
	\bnabla\bcdot \vel_j = 0,
\end{equation}
where  $\boldsymbol{u}_j$ and $p_j$ denote, respectively, the velocity and pressure fields in the fluid domains, with $j=i$ referring to the interior fluid ($a<r<b$) and $j=e$ referring to the exterior fluid ($r>b$).

The droplet interface is endowed with interfacial rheological properties in addition to surface tension. The mechanical balance of stresses across the interface requires that the traction jump between the exterior and interior fluids be balanced by capillary and other interfacial stresses. Denoting the traction jump by $\fD$, the dynamic boundary condition at the droplet interface $\GammaD$ can be written as 
\begin{equation}
\fD \equiv (\btau_e - \btau_i) \bcdot \unitNormal = 2 \gamma H \unitNormal  - \bnabla_s \bcdot \btaus,
\label{eq:TractionJump}
\end{equation}
where $\btau_j = -p_j \mathsfbi{I} + \mu_j (\bnabla \vel_j + \bnabla \vel_j^{\rm T})$ is the stress tensor in fluid $j$, $\mathsfbi{I}$ is the identity tensor, $\bnabla_s$ is the surface gradient operator, $\unitNormal$ is the outward unit normal, and $H = \frac{1}{2}\bnabla_s \bcdot \unitNormal$ is the mean curvature of the interface. We assume a uniform surface tension $\gamma$, thereby isolating the effects of interfacial rheology from Marangoni stresses. The interfacial stress tensor $\btaus$ is governed by an interfacial rheological constitutive model. In this work, we consider the Boussinesq-Scriven constitutive equation \citep{Scriven1960, Edwards1991}, 
\begin{equation}
    \btaus = 2\eta \mathsfbi{D}_s^{d} + \kappa \left( \bnabla_s \bcdot \uD \right) \mathsfbi{P},
    \label{eq:BoussinesqScrivenModel}
\end{equation}
to model the droplet interface with surface shear viscosity $\eta$ and surface dilatational viscosity $\kappa$. Here $\mathsfbi{P} = \mathsfbi{I} - \unitNormal\unitNormal^T$ is the surface projection tensor onto the tangent plane, and $\bnabla_s \bcdot \uD$ is the surface divergence of the velocity at the droplet interface, $\uD$. The surface rate of deformation tensor $\mathsfbi{D}_s$ is expressed in terms of the surface gradient of the velocity field,
\begin{equation}
\mathsfbi{D}_s = \frac{1}{2} \left[ \mathsfbi{P} \bnabla_s \uD + \left( \bnabla_s \uD \right)^{\rm T} \mathsfbi{P} \right],
\label{eq:SurfaceRateOfDeformation}
\end{equation}
and its deviatoric part is
\begin{equation}
\mathsfbi{D}_s^{d} = \mathsfbi{D}_s - \frac{1}{2} \left( \bnabla_s \bcdot \uD \right) \mathsfbi{P},
\label{eq:DeviatoricSurfaceRateOfDeformation}
\end{equation}
with the trace of $\mathsfbi{D}_s$ given by $\text{tr}(\mathsfbi{D}_s) = \bnabla_s \bcdot \uD$. The interfacial viscous force density $\boldsymbol{f}^V = \bnabla_s \bcdot \btaus$ comprises contributions from both surface shear and dilatational viscosities,
\begin{equation}
\boldsymbol{f}^V = \boldsymbol{f}^\eta + \boldsymbol{f}^\kappa,
\label{eq:ViscousForceDecomposition}
\end{equation}
where the contribution from surface shear viscosity is given by 
\begin{equation}
\boldsymbol{f}^\eta = 2 \eta \bnabla_s \bcdot \mathsfbi{D}_s^{d},
\label{eq:ShearViscousForce}
\end{equation}
and the contribution from surface dilatational viscosity is given by
\begin{equation}
\boldsymbol{f}^\kappa = \kappa \bnabla_s \bcdot \left[ \left( \bnabla_s \bcdot \uD \right) \mathsfbi{P} \right].
\label{eq:DilatationalViscousForce}
\end{equation}

We remark on several key dimensionless groups governing the dynamics of the problem. The confinement ratio, $\alpha = b/a$, measures the relative size of the droplet and the enclosed particle, and the viscosity contrast, $\lambda = \mu_i/\mu_e$, compares the viscosity of the fluid inside the droplet with that of the surrounding fluid. The capillary number, $Ca = \mu_e U_P/\gamma$, measures the relative importance of viscous stresses compared with surface tension.  Finally, we have the surface shear and dilatational Boussinesq numbers, $Bq_\eta = \eta/(\mu_e b)$ and $Bq_\kappa = \kappa/(\mu_e b)$, which characterize, respectively, the relative strength of interfacial shear and dilatational stresses as compared with the bulk viscous stress. Together, the dimensionless parameters $\alpha$, $\lambda$, $Ca$, $Bq_\eta$ and $Bq_\kappa$ capture the essential physical mechanisms governing the motion of the compound particle system. Throughout this work, we consider the limit of small capillary number ($Ca \ll 1$), where the droplet remains spherical. This approximation allows us to neglect shape deformation and isolate the influence of interfacial rheology on the motion of the compound particle system.

\section{Solution methods}

\subsection{Analytical solution}
\label{Sec:TheoreticalAnalysis}

We first consider the concentric configuration ($\epsilon = 0$) in which the rigid particle is located at the center of the droplet. We adopt spherical coordinates with the origin at the common center, where $r$ denotes the radial distance from the origin and the polar axis ($\theta=0$) is aligned with the axis of symmetry (the $z$–axis). In this geometry, the particle surface and the droplet interface correspond to the coordinate surfaces $r=a$ and $r=b$, respectively. The unit vectors in the radial and polar directions are denoted by $\boldsymbol{e}_r$ and $\boldsymbol{e}_\theta$, respectively. Owing to axisymmetry, the velocity field in fluid $j$, $\vel_j = u^r_j \boldsymbol{e}_r + u^\theta_j \boldsymbol{e}_\theta$, can be represented using a streamfunction formulation, where the velocity components are expressed in terms of a streamfunction $\psi_j$ as
\begin{equation}
	u_j^r = \frac{1}{r^2 \sin{\theta}} \frac{\partial \psi_j}{\partial \theta}, \qquad u_j^\theta = -\frac{1}{r \sin{\theta}} \frac{\partial \psi_j}{\partial r} \cdot
	\label{eq:VelocityComponentsFromStreamFunction}
\end{equation}
where $u^r_j$ and $u^\theta_j$ are, respectively, the radial and polar velocity components. The streamfunction formulation automatically satisfies incompressibility and reduces the Stokes equations to a biharmonic equation,
\begin{equation}
	E^4 \psi_j = 0,
	\label{eq:BiharmonicEquation}
\end{equation}
where $E^4 = E^2(E^2)$ and the operator $E^2$ is given by
\begin{equation}
	E^2 = \frac{\partial^2}{\partial r^2} + \frac{\sin{\theta}}{r^2} \frac{\partial}{\partial \theta} \left( \frac{1}{\sin{\theta}} \frac{\partial}{\partial \theta} \right).
	\label{eq:StokesOperator}
\end{equation}
The general solutions for the streamfunctions in the exterior and interior fluid domains take the form \citep{happel2012low}
\begin{align}
\psi_e(r, \theta) &= \sin^2{\theta} \left( \frac{1}{10} C_1 r^4 + C_2 r^2 - \frac{1}{2} C_3 r + \frac{C_4}{r} \right) \quad \text{for} \quad r \ge b,
\label{eq:StreamFunctionExterior} \\
\psi_i(r, \theta) &= \sin^2{\theta} \left( \frac{1}{10} C_5 r^4 + C_6 r^2 - \frac{1}{2} C_7 r + \frac{C_8}{r} \right) \quad \text{for} \quad a \le r \le b.
\label{eq:StreamFunctionInterior}
\end{align}
The unknown coefficients $C_i$ ($i=1\ldots8$), together with the droplet velocity \(U_D\), are determined by imposing boundary conditions at the rigid particle surface and droplet interface, along with appropriate far-field conditions. 

Working in the laboratory frame, the flow in the far-field is quiescent, 
\begin{align}
\vel_e(r\to\infty, \theta) \to \boldsymbol{0}.
\end{align}
At the rigid particle surface ($r=a$), the impermeability and the no-slip condition require that the fluid velocity match the rigid-body translational velocity of the enclosed particle, 
\begin{align}
\vel_i(r=a,\theta) = U_P \boldsymbol{e}_z = U_P \cos \theta \boldsymbol{e}_r -U_P \sin\theta \boldsymbol{e}_\theta.
\end{align}
At the droplet interface ($r=b$), both the normal and tangential velocity components must be continuous across the interface, as the droplet translates at an unknown velocity, $\boldsymbol{U}_D =U_D \boldsymbol{e}_z$, to be determined as part of the solution. Continuity of the normal velocity requires 
\begin{align}
u_e^r(r=b,\theta) = u_i^r(r=b,\theta),
\end{align}
while continuity of the tangential velocity requires 
\begin{align}
u_e^\theta(r=b,\theta) = u_i^\theta(r=b,\theta).
\end{align}

The dynamic boundary condition at the droplet interface is given by the traction jump equation (\ref{eq:TractionJump}). In the small capillary number regime, the interface is assumed to remain spherical, and the normal stress balance is not used to determine the shape. The tangential stress balance at $r=b$ takes the form 
\begin{equation}
	\tau^{r\theta}_e(b, \theta) - \tau^{r\theta}_i(b, \theta) = -\frac{2 \kappa}{b^2}\frac{\partial u^r(b,\theta)}{\partial \theta} - \frac{\kappa + \eta}{b^2} \frac{\partial}{\partial \theta}\left\{\frac{1}{\sin{\theta}} \frac{\partial\left[\sin{\theta}u^\theta(b,\theta)\right]}{\partial \theta} \right\} -\frac{2 \eta u^\theta(b,\theta)}{b^2},
	\label{eq:TangentialStressBalance}
\end{equation}
where $\tau^{r\theta}_j$ is the $r\theta$--component of the stress tensor in fluid $j$ given by
\begin{equation}
\tau^{r\theta}_j(r, \theta) = \mu_j \left[ \frac{1}{r} \frac{\partial u_j^r}{\partial \theta} + r \frac{\partial}{\partial r} \left( \frac{u_j^\theta}{r} \right) \right].
\label{eq:StressTensorComponent}
\end{equation}
Finally, we impose an overall force balance requiring that the net hydrodynamic force on the particle ($r=a$) equals that on the enclosing drop ($r=b$). This condition provides the final equation needed to close the system. Together with the boundary conditions at the particle surface and the droplet interface, it yields a linear algebraic system for the unknown coefficients $C_i$ and the droplet speed $U_D$. Upon solving this system, we obtain the following expressions for the coefficients,
\begin{equation}
C_1 = 0, \label{eqn:C1}
\end{equation}
\begin{equation}
C_2 = 0,
\end{equation}
\begin{equation}
	C_3 = -\frac{3 a \alpha \lambda \left[ \left( \lambda - 1 \right) \left( 2 + 3 \alpha^5 \right) + 5 \alpha^5 + 2 \left( \alpha^5 -1 \right) Bq_{\kappa}\right]  U_P}{\mathcal{C}},
\label{eq:CoeffB}
\end{equation}
\begin{equation}
	C_4 = -\frac{a^3 \alpha^3 \lambda \left[ \left( \lambda - 1 \right) \left( 2 + 3 \alpha^5 \right) + 5 \alpha^3 + 2 \left( \alpha^5 -1 \right) Bq_{\kappa}\right]  U_P}{2\mathcal{C}},
\label{eq:CoeffD}
\end{equation}
\begin{equation}
	C_5 = - \frac{15 \alpha \left( \alpha^2 - 1 \right) \left( \lambda - 1 - Bq_\kappa \right)  U_P}{a^2 \mathcal{C}},
\label{eq:CoeffE}
\end{equation}
\begin{align}
	C_6 = \frac{1}{2 \mathcal{C}}\bigg\{(\lambda-1)\left[ 2(\lambda-1) (2+3\alpha^5)+5\alpha^3(3\alpha^2-1) \right] \notag \\  
    + \left[ 4(\lambda-1)(\alpha^5-1)-5\alpha^3 (\alpha^2-1) \right] Bq_\kappa\bigg\} U_P,
\label{eq:CoeffG}
\end{align}

\begin{equation}
	C_7 = \frac{C_3}{\lambda},
\label{eq:CoeffF}
\end{equation}
\begin{equation}
	C_8 = -\frac{a^3 \alpha^3  \left[ \left( \lambda - 1 \right) \left( 2 + 3 \alpha^3 \right) + 5 \alpha^3 + 2 \left( \alpha^3 - 1 \right) Bq_{\kappa}\right]  U_P}{2\mathcal{C}},
\label{eqn:C8}
\end{equation}
together with the speed of the droplet,
\begin{equation}
    U_D = \frac{\lambda \left[ 2 \left( \lambda - 1 \right) \left( 2 + 3 \alpha^5 \right) + 5 \alpha^{3} \left( 3 \alpha^2 - 1 \right)  + 4 \left( \alpha^{5} - 1 \right) Bq_\kappa \right] U_P}{\mathcal{C}},	\label{eq:DropletVelocity}
\end{equation}
where 
\begin{align}
	\mathcal{C} = \left( \lambda - 1 \right) \left[ 4 \left( \lambda - 1 \right) + 9 \alpha - 10 \alpha^{3} + 3 \left( 2 \lambda + 3 \right) \alpha^{5} + 6 \alpha^{6} \right] + 10 \alpha^{6} \notag \\
    + \bigg\{4 \left( \alpha^{5} - 1 \right) \lambda + \left( \alpha - 1 \right)^{4} \left[ 4 + \alpha \left( 7 + 4 \alpha \right) \right] \bigg\} Bq_\kappa.
\label{eq:CDefinition}
\end{align}
The analytical solution given by (\ref{eq:DropletVelocity}) reveals that the induced droplet speed $U_D$ is independent of the shear Boussinesq number $Bq_\eta$, \textit{i.e.,} it is unaffected by surface shear viscosity. This  feature has also been observed for the motion of a translating droplet with surface viscosities \citep{Narsimhan2018} and for the motion of such a droplet enclosing a concentric squirmer \citep{Nganguia2025}. We further discuss this result for the concentric configuration and examine numerically whether this qualitative feature persists in eccentric cases in \S\ref{Sec:ResultsAndDiscussion}.

In addition to the analytical solution for the concentric configuration above, we derive in Appendix \ref{Sec:AppendixB} an integral relation based on the reciprocal theorem \citep{happel2012low, masoud_stone_2019} for the more general problem of a translating spherical object driven by an external force, applicable to both concentric and eccentric configurations. This integral relation is also used to validate the numerical implementation of the boundary integral method discussed below.

\subsection{Boundary integral method}
\label{Sec:BoundaryIntegralMethod}

To extend the investigation to eccentric particle configurations, we adopt a three-dimensional computational framework based on a spectral boundary integral method \citep{Zhao2010, Graham2002, Chao2021, Gurbuz2021, Rahimian2015}. Exploiting the linearity of the Stokes equations, this formulation recasts the governing equations as boundary integral equations posed solely on the bounding surfaces \citep{Youngren1975, Pozrikidis1992, Pozrikidis2001, BarthesBiesel2016, Atkinson1997}, thereby reducing the three-dimensional problem to a set of coupled two-dimensional surface integrals that naturally accommodate arbitrary particle positions.

The integral formulation involves the Stokeslet $\Gbold$ and the stresslet $\Kbold$, which are fundamental solutions of Stokes flow corresponding to a point force and a point-force dipole, respectively \citep{Ladyzhenskaya1969, KimKarrila, Pozrikidis1992}:
\begin{equation}
\Gbold(\chibold, \xibold) = \frac{1}{\abs{\rbold}} \left( \mathsfbi{I} + \hat{\rbold} \hat{\rbold}^T \right), \qquad
\Kbold(\chibold, \xibold) = \frac{6}{\abs{\rbold}^2} \left[ \hat{\rbold}^T \unitNormal(\xibold) \right] \hat{\rbold} \hat{\rbold}^T,
\label{eq:KernelFunctions}
\end{equation}
where $\rbold = \chibold - \xibold$ denotes the vector from the field point $\xibold$ to the target point $\chibold$, $\abs{\rbold}$ its magnitude, and $\hat{\rbold} = \rbold/\abs{\rbold}$ the corresponding unit vector. When the target point $\chibold$ lies on the droplet interface, the boundary integral equation takes the form
\begin{multline}
    \uD(\chibold)
    + \frac{\lambda - 1}{8 \pi} \int_{\GammaD} \Kbold(\chibold, \xibold) \left[ \uD(\xibold) - \uD(\chibold) \right] \dGamma 
    = \uP_e(\chibold)
    - \frac{1}{8 \pi \mu_e} \int_{\GammaD} \Gbold(\chibold, \xibold) \fD(\xibold) \dGamma,
	\quad \chibold \in \GammaD,
    \label{eq:BoundaryIntegralEquation1}
\end{multline}
where $\uP_e(\chibold)$ denotes the exterior velocity field due to the translating rigid particle, evaluated at the droplet interface,
\begin{equation}
\uP_e(\chibold) = - \frac{1}{8 \pi \mu_e} \int_{\GammaP} \Gbold(\chibold, \xibold) \fP(\xibold) \dGamma,
\label{eq:ParticleExteriorVelocity}
\end{equation}
and $\fP$ is the traction field on the surface of the enclosed particle. This expression corresponds to the standard boundary integral representation for the velocity field around a rigid particle translating in an unbounded fluid of viscosity $\mu_e$.
In the compound particle system, this exterior velocity field is evaluated at the droplet interface to capture the hydrodynamic influence of the enclosed particle on the droplet. The left-hand side of (\ref{eq:BoundaryIntegralEquation1}) comprises the droplet surface velocity and a double-layer contribution arising from the viscosity contrast between the interior and exterior fluids \citep{Rahimian2015}, which vanishes when $\lambda = 1$.

When the target point lies on the particle surface, the corresponding boundary integral equation is given by
\begin{equation}
    \uP(\chibold) = \uD_i(\chibold)
    - \frac{1}{8 \pi \mu_i} \int_{\GammaP} \Gbold(\chibold, \xibold) \fP(\xibold) \dGamma,
	\quad \chibold \in \GammaP,
    \label{eq:BoundaryIntegralEquation2}
\end{equation}
where $\uD_i(\chibold)$ denotes the interior velocity field due to the droplet, evaluated at the particle surface:
\begin{multline}
    \uD_i(\chibold) = \frac{\lambda - 1}{\lambda} \left\{ \uD(\chibold_\perp) - \frac{1}{8 \pi} \int_{\GammaD} \Kbold(\chibold, \xibold) \left[ \uD(\xibold) - \uD(\chibold_\perp) \right] \dGamma \right\} 
    - \frac{1}{8 \pi \mu_i} \int_{\GammaD} \Gbold(\chibold, \xibold) \fD(\xibold) \dGamma.
\label{eq:DropletInteriorVelocity}
\end{multline}
This expression corresponds to the boundary integral representation of the velocity field within a droplet of viscosity $\mu_i$, comprising contributions from interfacial motion and the traction jump across the interface. In the compound particle system, this interior velocity field is evaluated on the particle surface to account for the hydrodynamic influence of the droplet on the enclosed particle. The quantity $\chibold_\perp$ denotes the point on the droplet interface $\GammaD$ closest to the target point $\chibold \in \GammaP$ and plays a role in regularizing the integrals as discussed below.

Both surfaces, $\GammaD$ and $\GammaP$, are represented as smooth mappings of the unit sphere, parameterized by the colatitude angle $\vartheta \in [0, \pi]$ and azimuthal angle $\varphi \in [0, 2\pi)$. The surface position vector is expanded in real spherical harmonics up to degree $N$,
\begin{equation}
\xibold(\vartheta, \varphi) = \sum_{n=0}^{N} \left[ \frac{a_{0n}^{\xibold}}{2}\mathcal{P}_n^0(\cos{\vartheta})
+ \sum_{m=1}^n \mathcal{P}_n^m(\cos{\vartheta})
\left(a_{mn}^{\xibold} \cos{m\varphi} - b_{mn}^{\xibold} \sin{m\varphi}\right) \right],
\label{eq:SphericalHarmonicsExpansion}
\end{equation}
where $\mathcal{P}_n^m$ denotes the normalized associated Legendre functions, and $a_{mn}^{\xibold}$ and $b_{mn}^{\xibold}$ are the real spectral coefficients of degree $n$ and order $m$. The velocity $\uD$, traction $\fP$, and other field variables are expanded similarly. Spherical harmonic transforms are computed using the SPHEREPACK library \citep{Adams1999, Swarztrauber2000, Swarztrauber2004}.
This spectral representation provides high accuracy for smooth solutions \citep{Orszag1974, Atkinson1982, Graham2002}. For the results presented herein, we use $N = 16$, corresponding to a grid of $(N+1) \times (2N+1)$ quadrature points on each surface.

The integrals in (\ref{eq:BoundaryIntegralEquation1})--(\ref{eq:DropletInteriorVelocity}) pose computational challenges due to the singular behavior of the kernels when the target point $\chibold$ lies on or near the integration surface. For weakly singular integrals, where the target point lies on the integration surface, we employ a regularization strategy based on the eigenfunction property of spherical harmonics \citep{Graham2002, Veerapaneni2011b, Rahimian2015}. Since spherical harmonics are eigenfunctions of the integral operator with the singular kernel on the unit sphere, the addition theorem \citep{Colton2012, Arfken2013} allows this kernel to be replaced by a smooth, spectrally convergent sum of unnormalized Legendre polynomials, $\sum_{n=0}^{N} \bar{\mathcal{P}}_n(\cos\zeta)$, where $\zeta$ is the great-circle angle between the target and field points. In addition, the spherical coordinate system is rotated so that its north pole is aligned with the target point \citep{Ganesh2004, Veerapaneni2011b}. With this rotation, the coordinate singularity at the pole, where $\sin\vartheta = 0$, removes the discontinuity introduced by the regularization from the integration domain. For nearly singular integrals, where the target point lies close to but not on the integration surface, the integrand exhibits a sharp peak near the closest point. This is addressed by identifying the closest point $\chibold_\perp$ on the integration surface and rotating the coordinate system to align the north pole with $\chibold_\perp$.
The coordinate singularity at the pole mitigates the integrand peak, improving quadrature accuracy without requiring adaptive refinement. 

The surface differential operators appearing in the interfacial rheology terms involve products of the velocity field with geometric quantities, which cause aliasing errors. To maintain accuracy, we employ a dealiasing procedure \citep{Zhao2010, Veerapaneni2011b}: the coefficient array is expanded to twice the original resolution ($2N$), the nonlinear operations are performed in physical space, and the result is transformed back and truncated to the original resolution ($N$). The coupled system (\ref{eq:BoundaryIntegralEquation1})--(\ref{eq:DropletInteriorVelocity}) is solved using a Galerkin formulation \citep{Rahimian2015, Chao2021, Gurbuz2021}. Rather than solving for velocity values at collocation points, we solve directly for the spherical harmonic coefficients of the unknown fields $\uD$ and $\fP$ using generalized minimal residual method (GMRES). The matrix-free formulation evaluates the boundary integrals on--the--fly, avoiding explicit assembly of the dense system matrix.

\begin{figure}
        \centering
        \includegraphics[width=1\textwidth]{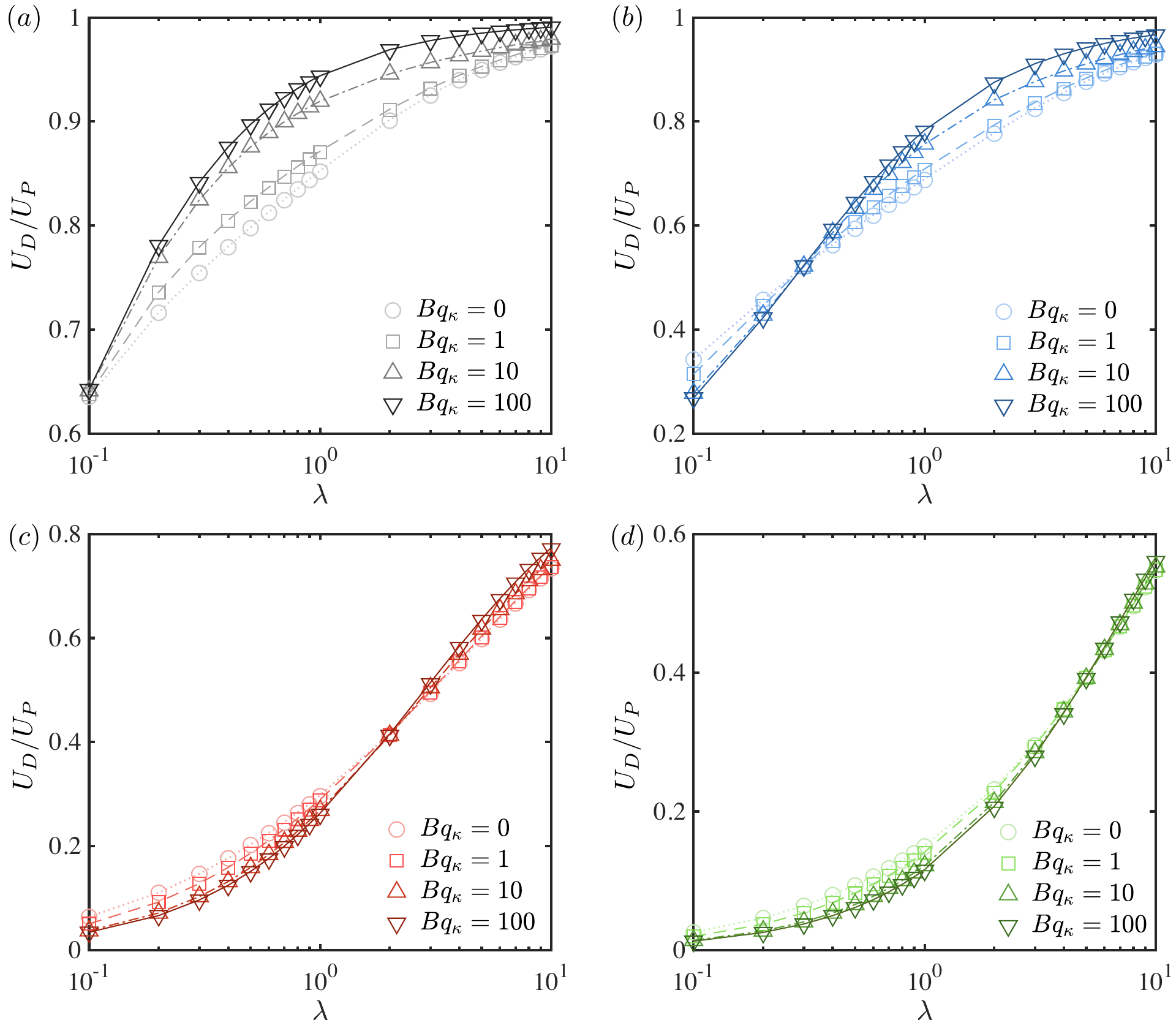}
    \caption{Induced droplet speed $U_D$, normalized by the prescribed particle speed $U_P$, as a function of the viscosity contrast $\lambda$ for various dilatational Boussinesq numbers $Bq_\kappa$ and confinement ratios: $(a)$ $\alpha = 1.5$, $(b)$ $\alpha = 2$, $(c)$ $\alpha = 5$, and $(d)$ $\alpha = 10$. The particle is concentric with the droplet ($\epsilon = 0$). Numerical results from the boundary integral method (symbols) are in excellent agreement with the analytical solution (lines).}
        \label{fig:UD_Lambda}
\end{figure}

\begin{figure}
        \centering
        \includegraphics[width=1\textwidth]{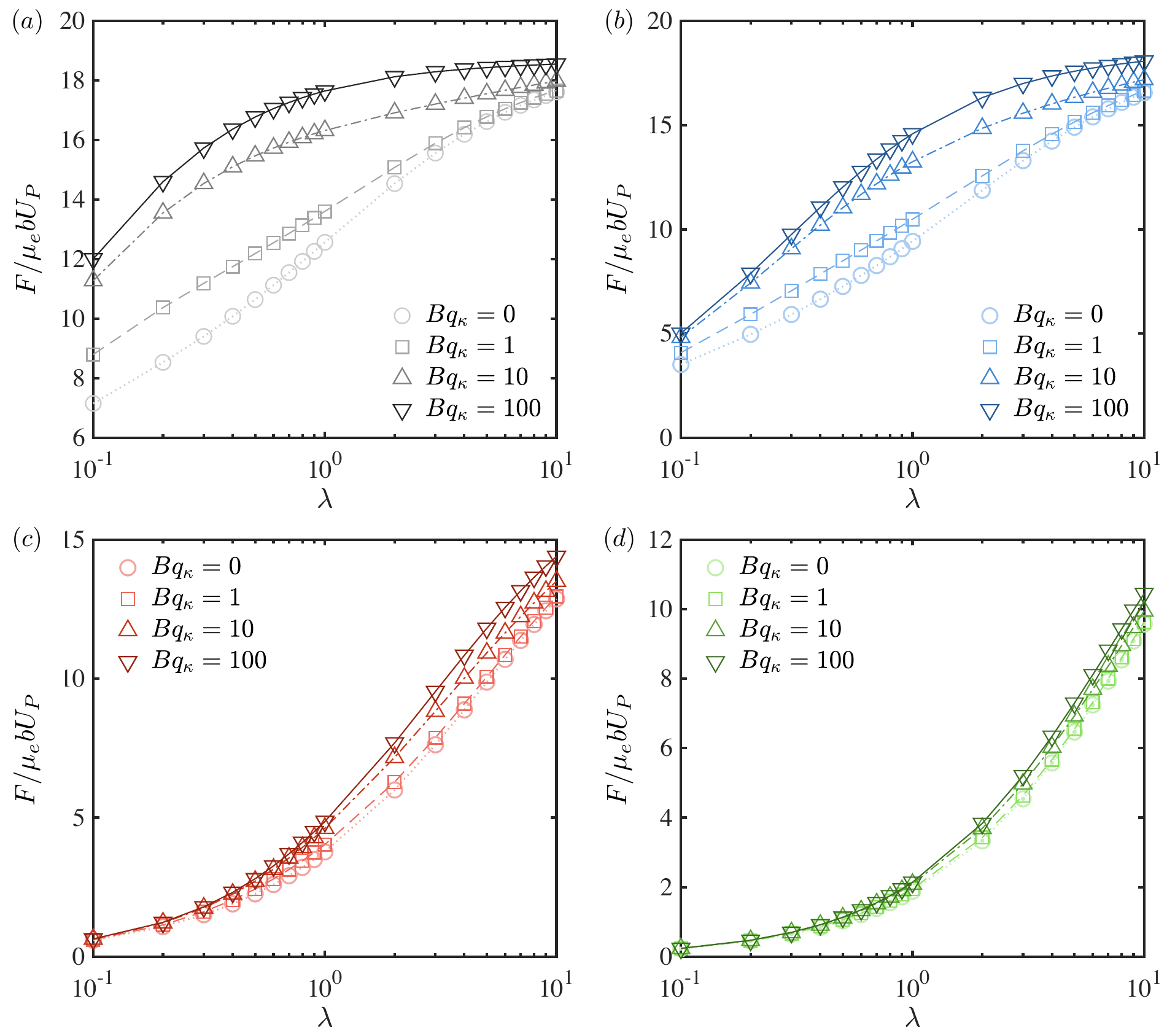}
    \caption{The magnitude of force $F$ required to drive the prescribed motion of the enclosed particle, normalized by the characteristic scale $\mu_e b U_P$, as a function of the viscosity contrast $\lambda$ for various dilatational Boussinesq numbers $Bq_\kappa$ and confinement ratios: $(a)$ $\alpha = 1.5$, $(b)$ $\alpha = 2$, $(c)$ $\alpha = 5$, and $(d)$ $\alpha = 10$. The particle is concentric with the droplet ($\epsilon = 0$). Numerical results from the boundary integral method (symbols) are in excellent agreement with the analytical solution (lines).}
        \label{fig:drag}
\end{figure}

\begin{figure}
        \centering
        \includegraphics[width=1\textwidth]{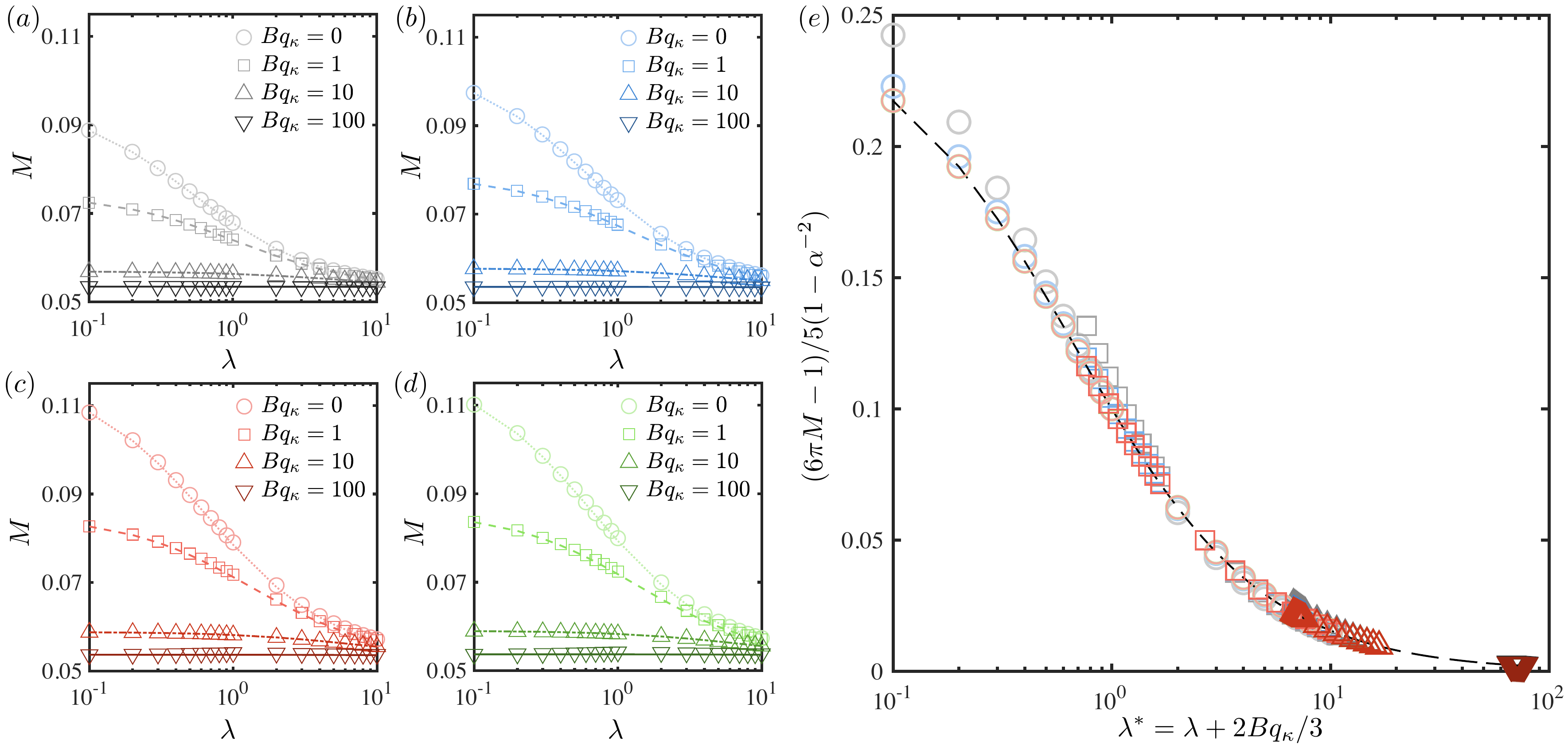}
    \caption{Droplet mobility $M$ as a function of the viscosity contrast $\lambda$ for various dilatational Boussinesq numbers $Bq_\kappa$ and confinement ratios: $(a)$ $\alpha = 1.5$, $(b)$ $\alpha = 2$, $(c)$ $\alpha = 5$, and $(d)$ $\alpha = 10$. $(e)$ Rescaled droplet mobility, $(6\pi M-1)/5(1-\alpha^{-2})$, plotted against the effective viscosity ratio, \(\lambda^* = \lambda + 2Bq_\kappa/3\). Data from panels $(a)$--$(d)$ largely collapse onto the dashed curve given by the asymptotic result, \((6\pi M-1)/5(1-\alpha^{-2})\sim 1/(4+6\lambda^*)\).}
        \label{fig:mobility}
\end{figure}

We validate the boundary integral method through two independent comparisons. First, for the concentric configuration, the numerical results are compared directly with the exact analytical solution derived in \S\ref{Sec:TheoreticalAnalysis}. As shown in figures \ref{fig:UD_Lambda}--\ref{fig:PostProcessing}, excellent agreement is observed across the full range of parameters considered. Second, we verify that the computed solutions satisfy a general integral relation derived from the Lorentz reciprocal theorem. This relation, which applies to both concentric and eccentric configurations, provides another validation that does not rely on the detailed analytical solution. The derivation and corresponding validation results are presented in Appendix \ref{Sec:AppendixB}.

\section{Results and discussion}
\label{Sec:ResultsAndDiscussion}
We present results based on the analytical solution and numerical simulations in this section, focusing first on the concentric configuration (\S\ref{Sec:ConcentricConfiguration}) before extending the discussion to eccentric cases (\S\ref{Sec:EccentricConfiguration}).

\subsection{Concentric configurations}
\label{Sec:ConcentricConfiguration}
For concentric configurations, the analytical result given by  (\ref{eq:DropletVelocity}), which is confirmed numerically, shows that the induced droplet speed $U_D$ is independent of the surface shear viscosity.
We therefore focus here on the effect of surface dilatational viscosity, characterized by the dilatational Boussinesq number $Bq_\kappa$, on the induced droplet speed. Figures~\ref{fig:UD_Lambda}$(a)$--$(d)$ show the results for the case of a clean droplet ($Bq_\kappa = Bq_\eta = 0$), indicated by circles (numerical results) and dotted lines (analytical results), serving as benchmarks for assessing the impact of increasing $Bq_\kappa$. The system exhibits qualitatively different behaviours depending on the confinement ratio $\alpha$ and the viscosity contrast $\lambda$. In a tightly confined system, for example with $\alpha = 1.5$, the presence of surface dilatational viscosity leads to an increase in $U_D$ across all values of $\lambda$ shown in figure~\ref{fig:UD_Lambda}$(a)$. When the particle is less confined, for example with $\alpha = 2$ and $5$, whether surface dilatational viscosity increases or decreases $U_D$ depends on $\lambda$. As shown in figures~\ref{fig:UD_Lambda}$(b)$ and \ref{fig:UD_Lambda}$(c)$, there exist critical viscosity ratios beyond which increasing $Bq_\kappa$ enhances $U_D$; below these thresholds, $U_D$ decreases with increasing $Bq_\kappa$. In the weak confinement regime where $\alpha = 10$, surface dilatational viscosity generally reduces $U_D$ across most values of $\lambda$, as illustrated in figure~\ref{fig:UD_Lambda}$(d)$.

To better understand the qualitatively different behaviours observed, we examine the interplay between interfacial rheology and the specifics of the problem setup. As discussed in previous studies \citep{Narsimhan2018, LEVAN198111, Singh2020, Dandekar_Ardekani_2020}, an increase in surface dilatational viscosity reduces interfacial mobility by introducing resistance to surface deformation. This added resistance tends to impede droplet motion. Accordingly, under a fixed driving force, one would expect that increasing $Bq_\kappa$ leads to a reduction in the droplet speed $U_D$. However, a key feature of the present setup is that the speed of the enclosed driving particle is held fixed. As a result, the force required to maintain the prescribed speed $U_P$ varies with the interfacial conditions. Specifically, reduced interfacial mobility due to increasing $Bq_\kappa$ necessitates a larger driving force to sustain the same particle speed. This force is transmitted to the surrounding fluid and, in turn, acts as the force inducing the motion of the enclosing droplet.

We therefore attribute the complex trends observed in figure~\ref{fig:UD_Lambda} to the competing effects of (i) reduced droplet mobility due to interfacial resistance and (ii) increased driving force arising from the fixed-speed constraint on the enclosed particle. To illustrate this competition, we present in figure~\ref{fig:drag} the magnitude of the total hydrodynamic force $F$ acting on the enclosed particle, which is equal to the force transmitted to the droplet.
The results show that increasing $Bq_\kappa$ generally leads to a greater force required to drive the motion of the enclosed particle at the same speed. This effect is particularly pronounced under tight confinement with $\alpha = 1.5$, as shown in figure~\ref{fig:drag}$(a)$, and less so under weak confinement with $\alpha=10$, as shown in figure~\ref{fig:drag}$(d)$. Consequently, for $\alpha = 1.5$, the dominant effect is the increased driving force, leading to a net increase in $U_D$, as seen in figure~\ref{fig:UD_Lambda}$(a)$. In contrast, for $\alpha = 10$, the reduction in droplet mobility dominates, resulting in a decrease in $U_D$, as shown in figure~\ref{fig:UD_Lambda}$(d)$. For intermediate confinement with $\alpha = 2$ and $\alpha = 5$, the two effects compete in a manner that depends on the viscosity ratio $\lambda$, giving rise to the more complex behaviours shown in figures~\ref{fig:UD_Lambda}$(b)$ and $(c)$.

To further illustrate the reduction in droplet mobility discussed above, we examine the relation between the droplet speed, $U_D = MF/(\mu_e b)$, and the applied external force $F$, through the mobility coefficient $M$ given by
\begin{align}
M = \frac{1}{6\pi} \left\{ 1 + \frac{5 \alpha^3 (\alpha^2-1)}{2\left[ \left( \lambda - 1 \right) \left( 2 + 3 \alpha^5 \right) + 5 \alpha^5 + 2 \left( \alpha^5 -1 \right) Bq_{\kappa}\right]} \right\}, \label{eq:mobility}
\end{align}
which characterizes the mobility of the droplet. From (\ref{eq:mobility}), $M$ decreases monotonically with increasing $Bq_\kappa$, as shown in figures \ref{fig:mobility}$(a)$--$(d)$ across all viscosity ratios $\lambda$ and confinement ratios $\alpha$. In the limit $Bq_\kappa \rightarrow \infty$, $M \rightarrow 1/6\pi$, approaching the mobility of a hard sphere. Further insights may be gained by rewriting the mobility coefficient as
\begin{align}
\frac{6\pi M-1}{5(1-\alpha^{-2})} = \frac{1}{2\left[2 (\lambda^{**}-1)\alpha^{-5}+(2+3\lambda^*) \right]}, \label{eq:mobilityScaled}
\end{align}
where $\lambda^{*} = \lambda + 2Bq_\kappa/3$ and $\lambda^{**}= \lambda - Bq_\kappa$. The factor $\lambda^*$ was noted previously for a simple translating drop \citep{Narsimhan2018, LEVAN198111}, where $Bq_\kappa$ acts as an effective increase in the interior viscosity. In the present problem, an additional dependence arises through $\lambda^{**}$, where increasing $Bq_\kappa$ may be interpreted as reducing the effective interior viscosity through this term. This latter effect, however, is less significant compared with the former effect due to $\lambda^*$, particularly in weakly confined systems with large $\alpha$, as it is scaled by $\alpha^{-5}$.

\begin{figure}
        \centering
        \includegraphics[width=1\textwidth]{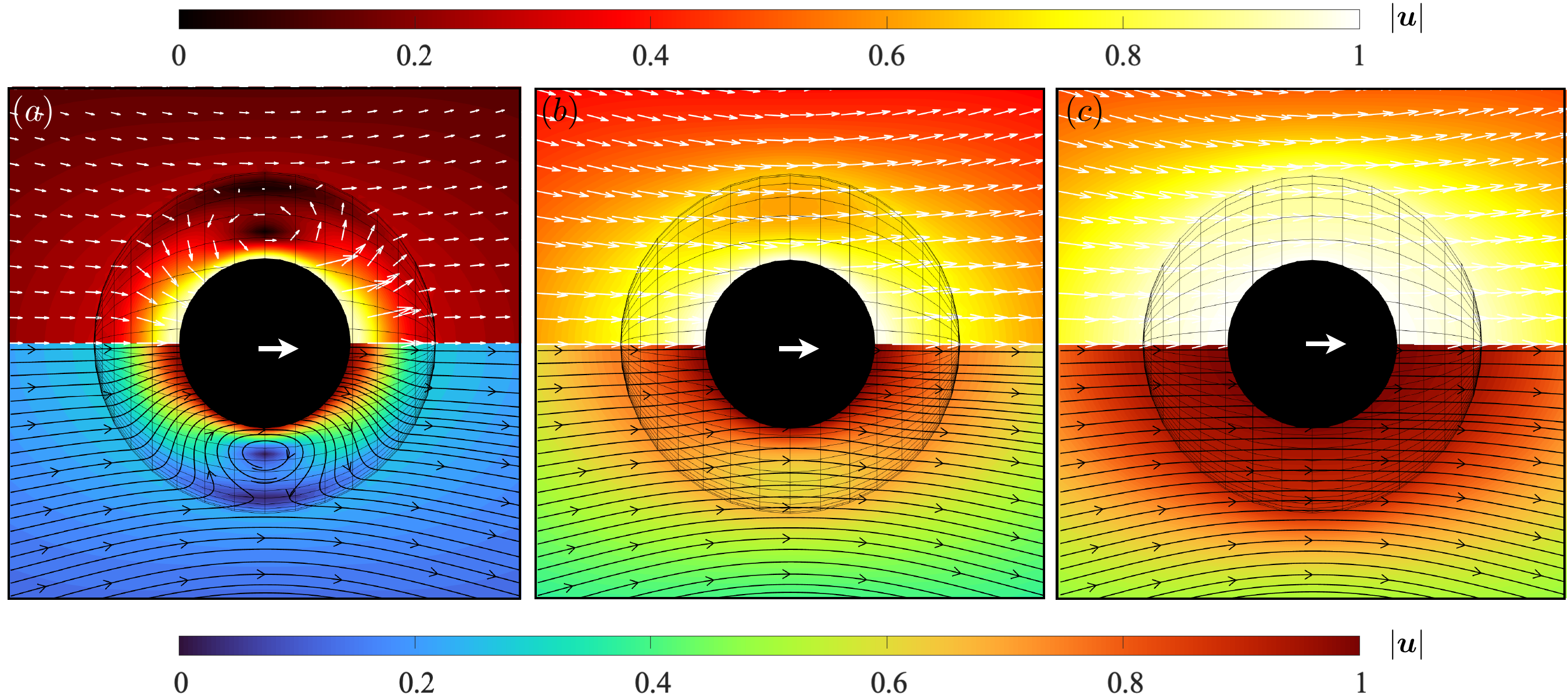}
\caption{Flow field inside and outside a viscous droplet enclosing a translating spherical particle in the concentric ($\epsilon = 0$) configuration for different viscosity contrasts: $(a)$ $\lambda =0.1$, $(b)$ $\lambda =1$, and $(c)$ $\lambda = 10$. Here, the Boussinesq numbers $Bq_\kappa = Bq_\eta = 10$ and confinement ratio $\alpha = 2$. In each panel, the upper half shows the numerical solution obtained using the boundary integral method, while the lower half shows the analytical solution. The colormap indicates the magnitude of the fluid velocity, $|\boldsymbol{u}|$. The white arrow on the black enclosed particle indicates its prescribed direction of motion.}
    \label{fig:PostProcessing}
\end{figure}

In the large--$\alpha$ regime, the right hand side of  (\ref{eq:mobilityScaled}) asymptotes to $1/(4+6\lambda^*)$, which is shown as the dashed curve in figure \ref{fig:mobility}$(e)$. Results in figures \ref{fig:mobility}$(a)$--$(d)$ largely collapse onto this curve, with only small deviations at lower $\lambda^*$, attributable to the contribution from the $\lambda^{**}$ term. Overall, these results suggest that the reduction in droplet mobility with increasing $Bq_\kappa$ may be interpreted primarily as an effective increase in the interior viscosity, particularly when $\lambda$ or $\alpha$ is large.

In figure \ref{fig:PostProcessing}, we compare the flow inside and outside the droplet obtained from the boundary integral method (upper halves) and the analytical solution (lower halves) for different viscosity ratios, showing excellent agreement. The overall flow characteristics are similar to those of the clean case \citep{KVS2019, Kawakami_Vlahovska_2025, Sprenger2020}. For a low viscosity ratio, for example $\lambda = 0.1$, the translating particle induces a recirculating flow inside the drop, as shown in figure \ref{fig:PostProcessing}$(a)$. This recirculation disappears as the viscosity ratio increases to $\lambda = 1$ and 10, as shown in figures \ref{fig:PostProcessing}$(b)$ and $(c)$.

\begin{figure}
        \centering
        \includegraphics[width=1\textwidth]{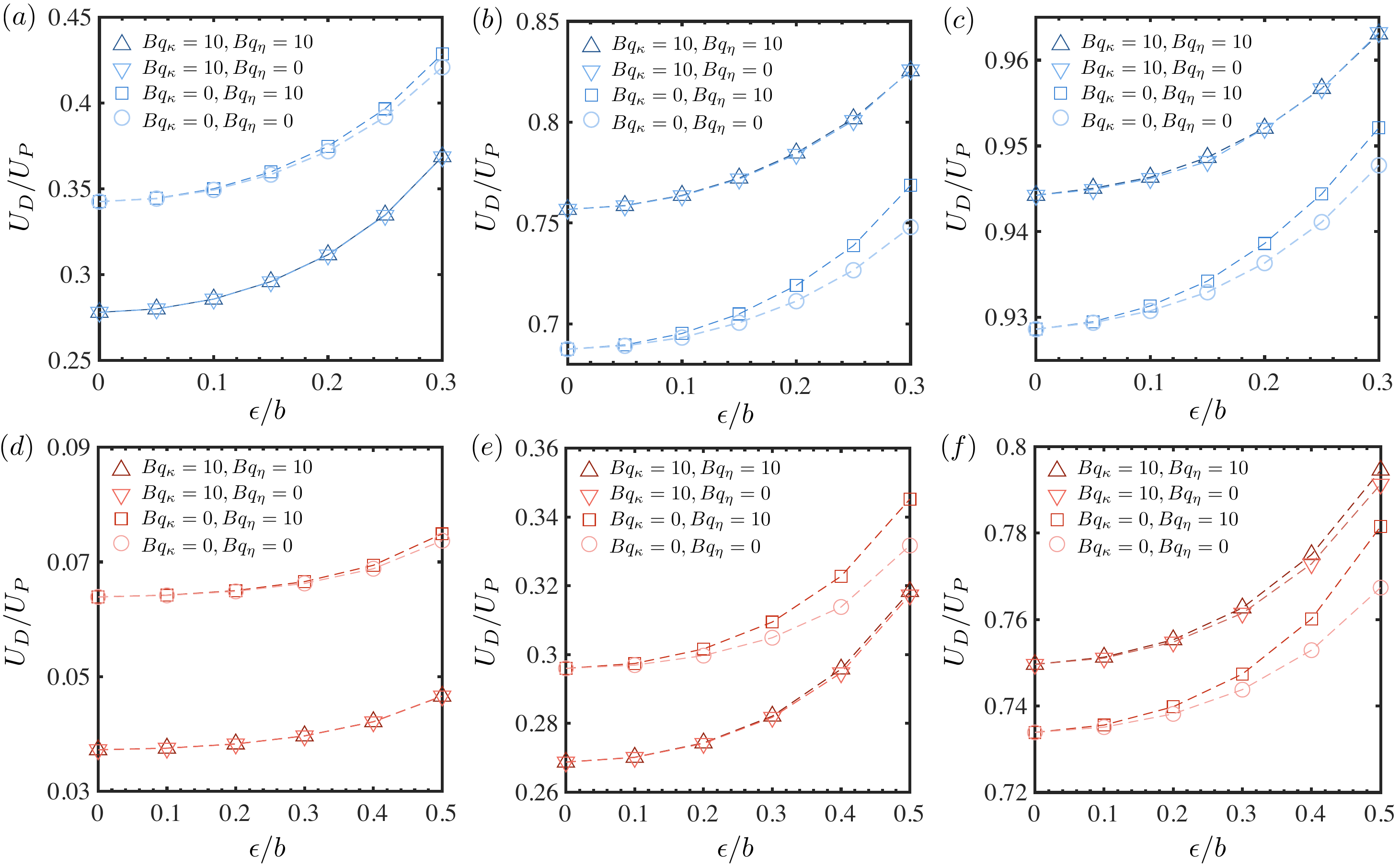}
    \caption{Induced droplet speed $U_D$, normalized by the prescribed particle speed $U_P$, as a function of the scaled eccentricity $\epsilon/b$ for various combinations of the Boussinesq numbers $Bq_\kappa$ and $Bq_\eta$ (see legends). The upper panels $(a)$--$(c)$ correspond to a confinement ratio of $\alpha = 2$ with viscosity ratios $\lambda = 0.1$, $1$, and $10$, respectively. The lower panels $(d)$--$(f)$ correspond to a confinement ratio of $\alpha = 5$ with viscosity ratios $\lambda = 0.1$, $1$, and $10$, respectively.}
    \label{fig:Eccentric}
\end{figure}

\begin{figure}
        \centering
        \includegraphics[width=1\textwidth]{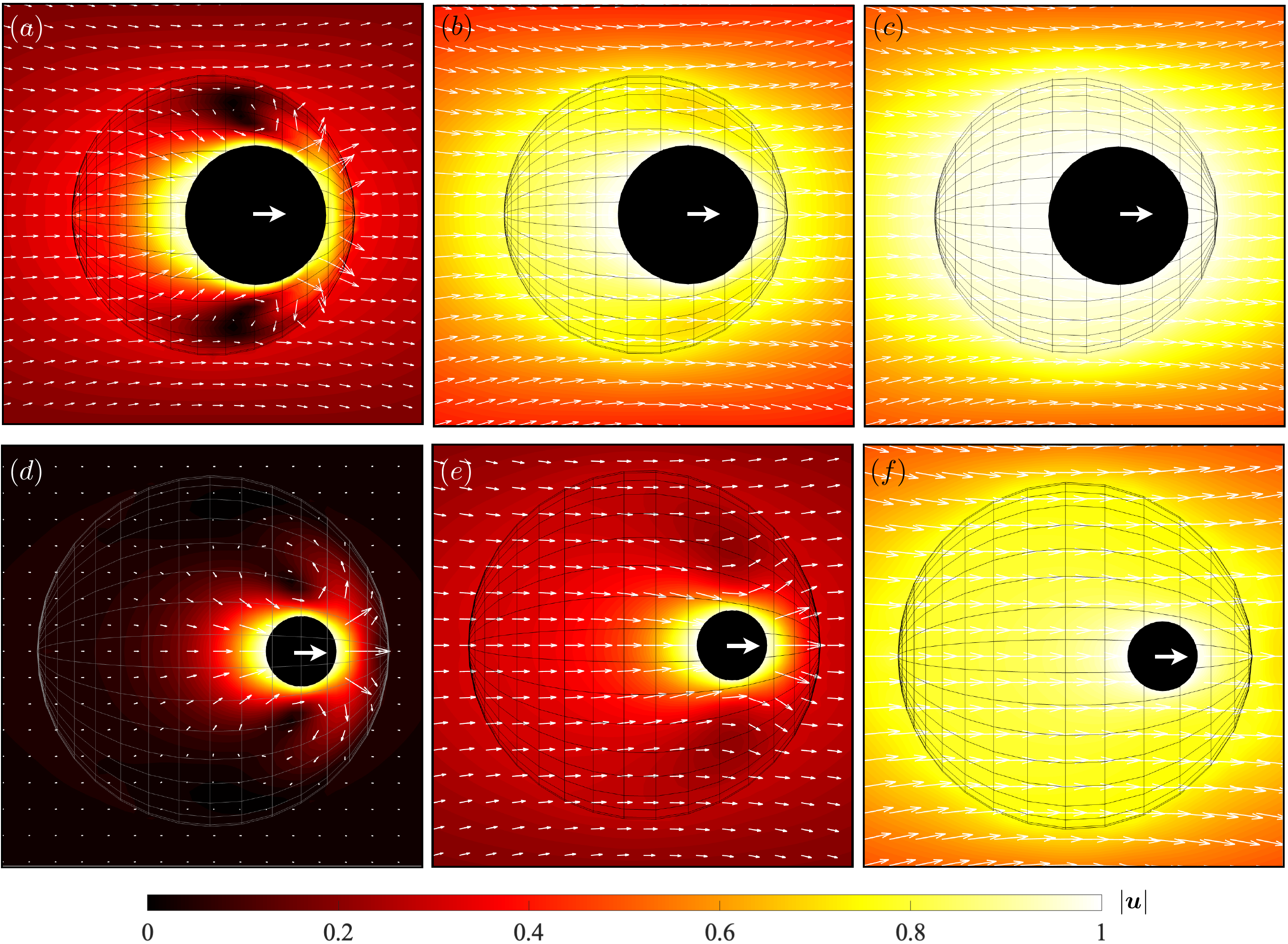}
\caption{Flow field inside and outside a viscous droplet enclosing a translating spherical particle in an eccentric configuration ($\epsilon/b = 0.3$), shown for different viscosity contrasts and confinement ratios. The upper panels $(a)$--$(c)$ correspond to a confinement ratio of $\alpha = 2$ with viscosity ratios $\lambda = 0.1$, $1$, and $10$, respectively. The lower panels $(d)$--$(f)$ correspond to $\alpha = 5$ with the same set of viscosity contrasts. Here, the Boussinesq numbers $Bq_\kappa = Bq_\eta = 10$. The colormap indicates the magnitude of the fluid velocity, $|\boldsymbol{u}|$. The white arrow on the black enclosed particle indicates its prescribed direction of motion.}
    \label{fig:PostProc_Eccentric}
\end{figure}

\subsection{Eccentric configurations}
\label{Sec:EccentricConfiguration}

We employ the boundary integral method to investigate eccentric  configurations, where the particle center is displaced from the droplet center. The eccentricity $\epsilon$ is defined as the distance between the particle and droplet centers (see figure \ref{fig:SchematicForCompoundDropletProblem}). 
In the concentric case ($\epsilon=0$), results in \S\ref{Sec:ConcentricConfiguration} showed that the droplet motion depends solely on surface dilatational viscosity and is unaffected by surface shear viscosity. Here, we assess whether these features continue to hold when the particle is off-center, examining the influence of both surface dilatational and shear viscosities.

As shown in figure~\ref{fig:Eccentric}, the induced droplet speed $U_D$ generally increases with eccentricity $\epsilon$ for different viscosity ratios, regardless of whether or not surface viscosities are present. We first isolate the effect of surface dilatational viscosity by comparing the clean case ($Bq_\kappa = Bq_\eta = 0$) with the case when only surface dilatational viscosity is present ($Bq_\kappa = 10$ and $Bq_\eta = 0$). The dependence of $U_D$ on $Bq_\kappa$ at various viscosity contrasts $\lambda$ and confinement ratios $\alpha$ mirrors that of the concentric case shown in figure~\ref{fig:UD_Lambda}. For example, for $\alpha=2$ and $\lambda=0.1$, shown in figure~\ref{fig:Eccentric}$(a)$, the presence of surface dilatational viscosity reduces $U_D$, indicated by downward triangles, compared with the clean case indicated by circles. However, as in the concentric case shown in figure~\ref{fig:UD_Lambda}$(b)$, increasing $\lambda$ to 1 or 10 reverses this trend, with $Bq_\kappa$ enhancing $U_D$ as shown in figures~\ref{fig:Eccentric}$(b)$ and $(c)$. For weaker confinement with $\alpha=5$, the transition from reduction to enhancement occurs at a higher critical $\lambda$, so that for $\lambda=1$, shown in figure~\ref{fig:Eccentric}$(e)$, $Bq_\kappa$ still decreases $U_D$, in contrast to the $\alpha=2$ case.

We next examine the effect of surface shear viscosity by comparing the clean interface case ($Bq_\kappa = Bq_\eta = 0$) with the case where only surface shear viscosity is present ($Bq_\kappa = 0$ and $Bq_\eta = 10$). In contrast to the concentric configuration, where $U_D$ is unaffected by surface shear viscosity, eccentric configurations exhibit a consistent increase in $U_D$ in the presence of surface shear viscosity across all cases in figure \ref{fig:Eccentric}, as evidenced by the comparison between circles (clean interface) and squares (interface with only surface shear viscosity). The influence of surface shear viscosity becomes more pronounced with increasing eccentricity $\epsilon$, and the associated speed enhancement is further amplified at larger viscosity contrasts $\lambda$. These results demonstrate that surface dilatational and shear viscosities affect droplet motion in qualitatively distinct ways. However, when both surface dilatational and shear viscosities are present (\textit{e.g.}, $Bq_\kappa = Bq_\eta = 10$), the droplet speed is largely determined by the dilatational contribution, and the additional effect of surface shear viscosity becomes comparatively weak, as shown by the comparison between downward (interface with only surface dilatational viscosity) and upward (interface with both surface dilatational and shear viscosities) triangles in figure \ref{fig:Eccentric}. Figure \ref{fig:PostProc_Eccentric} shows representative flow fields inside and outside the droplet for eccentric configurations. In contrast to the concentric cases shown in figure~\ref{fig:PostProcessing}, the interior flow exhibits clear fore–aft asymmetry about the droplet center. The velocity magnitude is higher in the narrow gap between the particle and the nearby interface, while the opposite side of the droplet exhibits weaker flow. This asymmetry is more pronounced at low viscosity contrast (\textit{e.g.}, $\lambda = 0.1$), as illustrated in figures \ref{fig:PostProc_Eccentric}$(a)$ and $(d)$, where distinct recirculation regions are observed. In these cases, the recirculation zones shift with the particle position and are located closer to the leading side of the interface. As in the concentric configurations, these recirculation regions are absent at higher viscosity contrasts (\textit{e.g.}, $\lambda = 1$ and $10$).

\section{Conclusion}

In this work, we have examined the hydrodynamics of a compound particle system with interfacial rheology, focusing on how surface viscosities influence the droplet motion induced by a translating particle enclosed within the droplet. An analytical solution was derived for the concentric configuration, and the analysis was extended to eccentric cases using boundary integral simulations. These approaches enabled a systematic quantification of the individual and combined effects of surface shear and dilatational viscosities on the induced droplet velocity across different configurations. 

For the concentric configuration, we find that the induced droplet velocity is independent of the surface shear viscosity. This behaviour may reflect a broader feature observed in related systems, including translating droplets with surface viscosities \citep{Narsimhan2018} and droplets enclosing concentric squirmers \citep{Nganguia2025}. In contrast, surface dilatational viscosity can enhance or reduce the induced droplet velocity depending on the confinement ratio and viscosity contrast. We interpret these trends in terms of two competing mechanisms, namely the increased force required to sustain the prescribed motion of the enclosed particle and the concurrent reduction in droplet mobility as surface dilatational viscosity increases. The balance between these effects gives rise to the observed non-monotonic dependence of droplet velocity on the surface dilatational viscosity. 

When the enclosed particle is eccentrically confined within the droplet, our results reveal an additional dependence on surface shear viscosity that is absent in the concentric case. In particular, surface shear viscosity consistently enhances the induced droplet velocity, with the effect becoming more pronounced as the eccentricity increases. However, when both surface dilatational and shear viscosities are present, the modification of the droplet velocity is determined primarily by surface dilatational contribution, with the influence of surface shear viscosity becoming less apparent. Taken together, our findings demonstrate that surface shear and dilatational viscosities can influence the induced motion of the compound droplet in qualitatively distinct ways. The results also underscore the role of geometric asymmetry in activating interfacial shear effects, providing insight into how interfacial rheology couples with confinement and symmetry breaking. 

We also remark on some assumptions and limitations in this work and discuss several future directions. First, the present numerical framework is not restricted to axisymmetric settings and can be extended to fully three-dimensional, non-axisymmetric configurations, including oblique particle motion, rotating or non-spherical inclusions, and multi-inclusion systems. Such scenarios are of practical interest in microfluidics, active matter, and complex encapsulated systems. In addition, in this work we restrict our attention to the non-deformable regime in which the droplet remains spherical. Investigations are underway to examine how droplet deformability interacts with interfacial rheological effects to further influence the dynamics of these compound particles. Finally, complex interfaces often exhibit not only viscous but also elastic effects \citep{Fuller2012, Jaensson2021}. While the present work focuses on interfacial viscous stresses, an important direction for future work is to examine how interfacial viscoelasticity influences the dynamics of the system. Overall, this work establishes a benchmark for future investigations, representing a first step toward exploring the dynamics of active compound particles with complex interfaces.\\

\noindent \textbf{Funding.} Y.N.Y.~acknowledges support from the National Science Foundation (DMS-1951600 and DMS-2510714) and Flatiron Institute, part of Simons Foundation. O.S.P.~acknowledges  support from the National Science Foundation (CBET-2323046 and CBET-2419945) and  the American Chemical Society Petroleum Research Fund (66541-ND9).\\ 

\noindent \textbf{Declaration of Interests.} The authors report no conflict of interest.

\appendix

\begin{figure}
        \centering
        \includegraphics[width=0.55\textwidth]{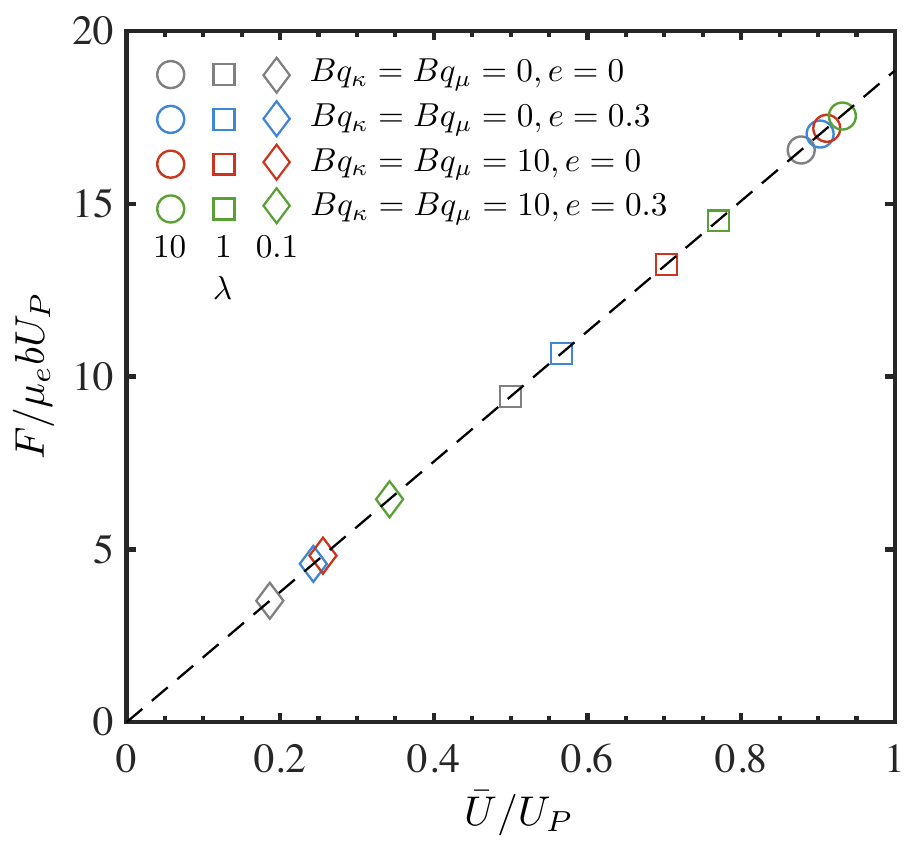}
    \caption{Scaled force, $F/\mu_e b U_P$, acting on the compound particle as a function of the scaled area-averaged surface velocity, $\bar{U}/U_P$. Numerical results obtained using the boundary integral method for both concentric ($\epsilon/b=0$) and eccentric ($\epsilon/b=0.3$) configurations, and for droplet interfaces without ($Bq_\kappa = Bq_\eta=0$) and with ($Bq_\kappa = Bq_\eta=10$) surface viscosities (symbols; see legend), collapse onto a single dashed line with slope $6\pi$, as predicted by the integral relation (\ref{eq:IntegralResult}).}
        \label{fig:IntegralRelation}
\end{figure}

\section{A general integral result for concentric and eccentric configurations}
\label{Sec:AppendixB}

In this appendix, we consider a more general problem setup consisting of a translating spherical object of radius $b$, corresponding to the outer boundary of the compound particle system in the main text, driven by a force $\boldsymbol{F}$ acting on the particle. The concentric and eccentric compound particle systems considered in the main text are therefore special cases of this general setup. 

Let $\vel$ and $\btau$ be the velocity and stress fields in this general problem. We utilize the reciprocal theorem \citep{happel2012low, masoud_stone_2019} to obtain an integral result relating the velocity $\vel$ on the surface of the translating spherical object  $\GammaD$ to the external force $\boldsymbol{F}$ acting on the object. To do so, consider an auxiliary Stokes flow problem where the flow velocity $\tilde{\vel}$ and stress $\tilde{\btau}$ fields denote the solution to the translation of the same spherical object at a velocity $\tilde{\boldsymbol{U}}$ subject to a force $\tilde{\boldsymbol{F}}$. The reciprocal theorem states that the solutions to the two problems are related by
\begin{align}
\int_{\GammaD} \unitNormal \bcdot \tilde{\btau}  \bcdot \vel \dif\Gamma = \int_{\GammaD} \unitNormal \bcdot \btau  \bcdot \tilde{\vel} \dif\Gamma. \label{eq:reciprocal}
\end{align}
In the auxiliary problem, $\tilde{\vel} = \tilde{\boldsymbol{U}}$ on $\GammaD$, so the right hand side of (\ref{eq:reciprocal}) becomes $\boldsymbol{F} \bcdot \tilde{\boldsymbol{U}}$, where $\boldsymbol{F} = \int_{\GammaD} \unitNormal \bcdot \btau  \dif\Gamma$ is the force acting on the particle. Furthermore, the surface traction in the auxiliary problem is given by, $\unitNormal \bcdot \tilde{\btau}  = -3\mu_e \tilde{\boldsymbol{U}}/2b$, for a spherical particle of radius $b$. Taken together, (\ref{eq:reciprocal}) can be rewritten as
\begin{align}
-\frac{3\mu_e}{2b}\tilde{\boldsymbol{U}} \bcdot \int_{\GammaD} \vel \dif\Gamma = \boldsymbol{F} \bcdot \tilde{\boldsymbol{U}}.
\end{align}
Considering an axisymmetric problem with $\boldsymbol{F}= F \boldsymbol{e}_z$ and a definition of an area average of the surface velocity,
\begin{align}
\bar{U} = \frac{1}{4\pi b^2} \left| \int_{\GammaD} \vel \dif\Gamma \right|, \label{eqn:Averaged}
\end{align}
the results in (\ref{eq:reciprocal}) can be expressed as
\begin{align}
F = 6 \pi \mu_e b \bar{U}. \label{eq:IntegralResult}
\end{align}
The integral relation in (\ref{eq:IntegralResult}) is valid regardless of the specific flow details within or on the surface of the spherical object, which determine the resulting velocity distribution on $\GammaD$. For the special case of a solid sphere translating at speed $U$, we have $\vel = U \boldsymbol{e}_z$ on $\GammaD$ and hence $\bar{U} = U$, reducing (\ref{eq:IntegralResult}) to the simple classical Stokes' law, 
\begin{align}
F = 6\pi \mu_e b U.
\end{align}
For another special case of a spherical drop translating at speed $U$, we have $\vel = U \boldsymbol{e}_z + U\sin\theta/2(1+\lambda) \boldsymbol{e}_\theta$ on $\GammaD$ and hence $\bar{U} = (3\lambda+2)U/3(1+\lambda)$, reducing to the Hadamard-Rybczynski solution \citep{Hadamard1911, Rybczynski1911}, 
\begin{align}
F =  \frac{2\pi \mu_e b U(3\lambda +2)}{\lambda+1} \cdot
\end{align}
For the compound particle system considered in the main text, the velocity $\vel$ on $\GammaD$ corresponds to the droplet surface velocity $\uD$. For the concentric configuration, the analytical solution derived in \S\ref{Sec:TheoreticalAnalysis} yields
\begin{align}
\uD = U_D \cos\theta \, \boldsymbol{e}_r + \frac{1}{b}\left(\frac{C_3}{2} + \frac{C_4}{b^2}\right) \sin\theta \, \boldsymbol{e}_\theta,
\label{eq:DropletSurfaceVelocity}
\end{align}
where $C_3$ and $C_4$ are coefficients in the streamfunction given in (\ref{eq:CoeffB}) and (\ref{eq:CoeffD}). The area--averaged surface velocity defined in (\ref{eqn:Averaged}) is therefore given by $\bar{U} = -2C_3/(3b)$. The force on the compound particle follows from (\ref{eq:IntegralResult}) as
\begin{align}
F = -4\pi \mu_e C_3 = \frac{12\pi \mu_e b \, \lambda \left[ (\lambda - 1)(2 + 3\alpha^5) + 5\alpha^5 + 2(\alpha^5 - 1) Bq_\kappa \right] U_P}{\mathcal{C}} \cdot
\label{eq:ForceCompound}
\end{align}
We verify that this expression, derived from the integral relation (\ref{eq:IntegralResult}), is consistent with the force calculated by directly integrating the traction over the droplet surface using the analytical solution. We further use this integral relation to validate the boundary integral implementation for both concentric and eccentric configurations, as well as for droplet interfaces with and without surface viscosities. As shown in figure \ref{fig:IntegralRelation}, the numerically computed forces for all cases collapse onto the same dashed line predicted by the integral relation.

\bibliographystyle{unsrtnat}
\bibliography{references}

\end{document}